\documentclass[paper]{agujournal2019}
\usepackage{url}
\usepackage{lineno}
\usepackage{color}
\usepackage{booktabs}
\usepackage{amsmath}
\usepackage{amssymb}
\usepackage{graphicx}
\usepackage[export]{adjustbox}
\usepackage{diagbox}
\usepackage{multirow}
\draftfalse

\newcommand{\ssr}{   {Space Sci. Rev. }}

\newcommand{\jgr}{   {J. Geophys. Res.}}
\newcommand{\grl}{   {Geophys. Res. Lett.}}

\newcommand{\apjl}{   {Astrophys. J. Lett.}}
\newcommand{\prl}{   {Phys. Rev. Lett.}}

\newcommand{\aap}{   {Astronomy \& Astrophysics}}

\journalname{JGR: Space Physics}

\begin{document}


\title{Night-Side Relativistic Electron Precipitation Bursts in the Outer Radiation Belt: Insights from ELFIN and THEMIS}

\authors{Xi Lu \affil{1}, Xiao-Jia Zhang \affil{1},  Anton V. Artemyev \affil{2}, Vassilis Angelopoulos \affil{2}, Jacob Bortnik \affil{2} } 
\affiliation{1}{Department of Physics, University of Texas at Dallas, Richardson, Texas, USA}
\affiliation{2}{University of California, Los Angeles, Los Angeles, California, USA}

\correspondingauthor{Xi Lu}{xi.lu@utdallas.edu}

\begin{keypoints}
\item We present a statistical study of unexpected relativistic electron precipitation bursts in the night-side outer radiation belt
\item We show that these precipitating bursts are likely driven by whistler-mode waves that propagate to middle latitudes
\item We parameterize the precipitating electron spectra for further applications in magnetosphere-ionosphere coupling models
\end{keypoints}

\begin{abstract}
Electromagnetic whistler-mode waves play a crucial role in the acceleration and precipitation of radiation belt electrons. Statistical surveys of wave characteristics suggest that these waves should preferentially scatter and precipitate relativistic electrons on the day side. However, the night-side region is expected to be primarily associated with electron acceleration. The recent low-altitude observations reveal relativistic electron precipitation in the night-side region. In this paper, we present statistical surveys of night-side relativistic electron losses due to intense precipitation bursts. We demonstrate that such bursts are associated with storm time substorm injections and are likely related to relativistic electron scattering by ducted whistler-mode waves. We also speculate on the role of injections in creating conditions favorable for relativistic electron precipitation.
\end{abstract}

\section{Introduction}
Electron resonant scattering by electromagnetic whistler-mode waves, particularly by the most intense whistler-mode chorus waves, is the primary driver of energetic electron losses in the outer radiation belt \cite{Bortnik&Thorne07,Millan&Thorne07,Li&Hudson19}. These wave-particle interactions result in electron pitch-angle scattering, leading to their precipitation into the Earth’s atmosphere where they may significantly alter ionospheric characteristics \cite{Seppala15, Mironova19, Verronen21:eep&aurora, Miyoshi21,Lyons21:frontiers,Nishimura21:agu}. 

The energy range of precipitating electrons is determined by two key characteristics of whistler-mode waves: the wave normal angle (WNA) and their latitudinal extent. 
Although the most intense whistler-mode waves are field-aligned chorus waves \cite{Li11,Agapitov13:jgr,Agapitov18:jgr}, these waves cannot resonate with relativistic electrons near the equatorial source region, where the energy of resonant electrons near the loss cone is limited to approximately $100-200$ keV \cite<e.g.,>{Summers07:rates,Summers07:theory,Ni11}. Therefore, to contribute to relativistic electron losses, whistler-mode waves should either be very oblique \cite{Lorentzen01,Mourenas14,Artemyev16:ssr,Gan23:grl_elfin} or propagate to middle latitudes away from the equator along the field lines, where resonance energies can exceed $500$ keV \cite<see discussions in>{Miyoshi20,Miyoshi21,Chen22:microbursts,Artemyev24:jgr:ELFIN&injection,Kang24:elfin}. Middle-latitude oblique waves, as well as field-aligned waves, are mainly observed in the day-side magnetosphere. In contrast, on the night side, whistler-mode waves are mostly confined around the equatorial region \cite{Bortnik07:model, Agapitov13:jgr,Agapitov18:jgr,Santolik14}. This day-night asymmetry in wave local time distribution, coupled with different ratios of plasma to cyclotron frequencies, indicates that day-side and night-side whistler-mode waves play different roles in radiation belt models: waves on the night-side are mostly implicated in the acceleration of electrons to relativistic energies, while those on the day-side contribute to both energetic and relativistic electron losses \cite{Meredith03,Horne05JGR,Mourenas14,Ma18,Agapitov19:fpe}. The recent low-altitude observations of electron precipitation suggest that night-side whistler-mode waves also contribute to relativistic electron losses \cite<see details in>{Shumko21,Tsai23,Tsai24}. These losses occur as intense bursts of precipitation, reaching the strong diffusion limit and typically occurring well equatorward from the curvature scattering region \cite<see examples in>{Artemyev24:jgr:ELFIN&injection,Kang24:elfin}. The mechanism underlying night-side relativistic precipitation is likely associated with a specific population of whistler-mode waves that can be ducted around plasma density gradients \cite{Hanzelka&Santolik19,Ke21:ducts,Hosseini21,Harid24}. These ducted waves can propagate to middle latitudes with high intensity \cite{Chen21:frontiers,Chen22:microbursts} and scatter relativistic electrons \cite{Miyoshi20,Kang24:elfin}. Although this relatively small population of ducted whistler-mode waves may not significantly impact average (statistical) wave models, they likely play a critical role in dictating night-side losses of relativistic electrons \cite<see discussions in>{Artemyev21:jgr:ducts,Artemyev24:jgr:ELFIN&injection,Tsai24}.

Observations of individual relativistic electron precipitation events  \cite{Douma17,Shumko21,Artemyev24:jgr:ELFIN&injection,Kang24:elfin} and their statistical contributions to electron losses \cite{Tsai23,Tsai24,Chen23:poes} highlight the necessity of including these losses in magnetosphere-ionosphere coupling models \cite<see the discussion in>{Zou24}. The significance lies in the fact that these precipitating electrons can drastically modify ionospheric properties below 100km \cite<e.g.,>{Seppala15, Oyama17:eep,  Yu18:substorm&precipitations, Stepanov21:eep, Verronen21:eep&aurora}. Therefore, it is crucial to investigate and parameterize the characteristics of this precipitation, including the energy spectra of precipitating electrons and their variability under different geomagnetic conditions.

In this work, we construct a dataset of relativistic precipitation bursts on the night side and investigate their statistical properties using high-resolution measurements from the ELFIN CubeSats \cite{Angelopoulos20:elfin}. These measurements provide detailed energy and full pitch-angle distributions. The identified bursts are detected equatorward from the curvature scattering region \cite<isotropy boundary; see details in>[and reference therein]{Wilkins23}, that is, well within the outer radiation belt where electron precipitation is primarily driven by whistler-mode waves. We fit the electron precipitation spectra with analytical functions and present the fitting parameters as functions of geomagnetic conditions and spatial location. These detailed characterizations of night-side precipitation can be readily incorporated into magnetosphere-ionosphere coupling models, improving their ability to predict ionospheric responses.

\section{Datasets}  
Two identical CubeSats from the Electron Losses and Fields Investigation (ELFIN) mission were launched on September 15, 2018 and ended on September 30, 2022 \cite{Angelopoulos20:elfin}. Each satellite is equipped with energetic particle detectors for ions (EPD-I) and electrons (EPD-E). Electrons are measured within an energy range of 50 keV to approximately $6$ MeV, with an energy resolution ($\Delta E/E$) of less than 40 \% and a time resolution ($T_{spin}$) of about $2.85$ s \cite{Angelopoulos23:ssr,Tsai24:review}. The high pitch-angle resolution of the EPD-E ($\Delta \alpha \simeq 22.5 ^{\circ}$) allows it to resolve the trapped electrons (outside the local bounce loss cone) from precipitating electrons (within the local bounce loss cone) \cite<see details>{Zhang22:microbursts,Angelopoulos23:ssr}. The ratio of the precipitating electron flux and the trapped electron flux ($J_{prec}/J_{trap}$) indicates the efficiency of equatorial electrons being scattered by waves \cite{Li13:POES,Mourenas21:jgr:ELFIN,Zhang22:natcom,Shen22:jgr:WISP,Capannolo23:elfin} or (and) by magnetic field line curvature \cite{Wilkins23,Zou24}.

Following previous case studies of energetic electron precipitation associated with whistler-mode waves \cite{Tsai22,Gan23:grl_elfin}, we examine individual bursts of electrons scattered by whistler-mode waves. The main criteria for distinguishing these bursts from those driven by electromagnetic ion cyclotron (EMIC) waves \cite{Angelopoulos23:ssr} and curvature scatterings \cite{Wilkins23} are as follows: [1] All events are observed equatorward from the electron isotropy boundary (IB) \cite<see details in>{Wilkins23,Artemyev23:ELFIN&dispersion,Sergeev23:elfin}, which is the inner edge of the curvature scattering region; [2] For all events, the ratio $J_{prec}/J_{trap}$ peaks at 50keV and decreases with increasing energy \cite< which differs from events associated with EMIC waves, see discussions in>{Grach22:elfin,Bashir24:jgr}, [3] All events exhibit $J_{prec}/J_{trap}>0.5$ at relativistic energies ($\sim 500$keV) \cite<see examples in>{Artemyev24:jgr:ELFIN&injection}. In total, 169 events of  relativistic electron precipitation bursts were detected on the night side ($MLT \in[20,5]$) from ELFIN observations in 2021 and 2022. 

Several night-side relativistic electron precipitation events captured by ELFIN are accompanied by near-equatorial observations from Time History of Events and Macroscale Interactions during Substorms (THEMIS) \cite{Angelopoulos08:ssr}. 
We use spin-resolution (FGS, $\sim3$ s) magnetic field data in geocentric solar magnetospheric (GSM) coordinates measured by the fluxgate magnetometer \cite{Auster08:THEMIS}. Additionally, we analyze the wave magnetic field spectra data from the FFF dataset during fast survey mode (1 s time resolution) \cite{Cully08:ssr}, utilizing search-coil magnetometer measurements \cite{LeContel08}. Electron density data is provided by the electrostatic analyzer (ESA) in the reduced mode ($\sim3$ s time resolution) \cite{McFadden08:THEMIS}.

\begin{figure*}
\centering
\includegraphics[width=1\textwidth]{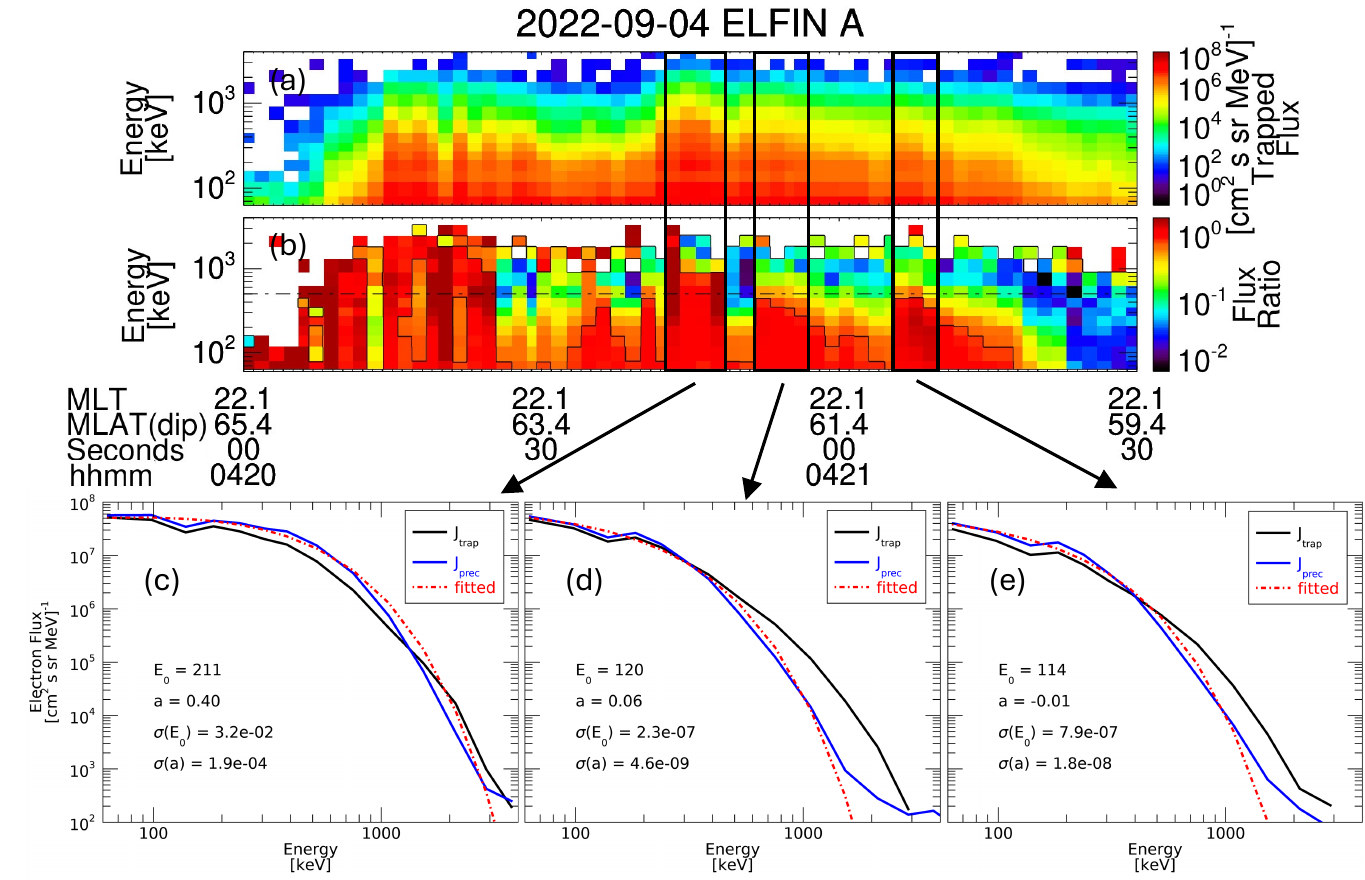}
\caption{First example of night-side ELFIN observations of relativistic electron precipitation bursts from the outer radiation belt: (a) locally trapped electron fluxes, (b) precipitating-to-trapped flux ratio with precipitating bursts marked by black rectangles, (c)-(e) trapped and precipitating electron spectra within the relativistic bursts, with the precipitating spectrum fitted by $A\cdot\left(E/E_0\right)^a\cdot\exp(-E/E_0)$.
\label{fig1}}
\end{figure*}

\begin{figure*}
\centering
\includegraphics[width=1\textwidth]{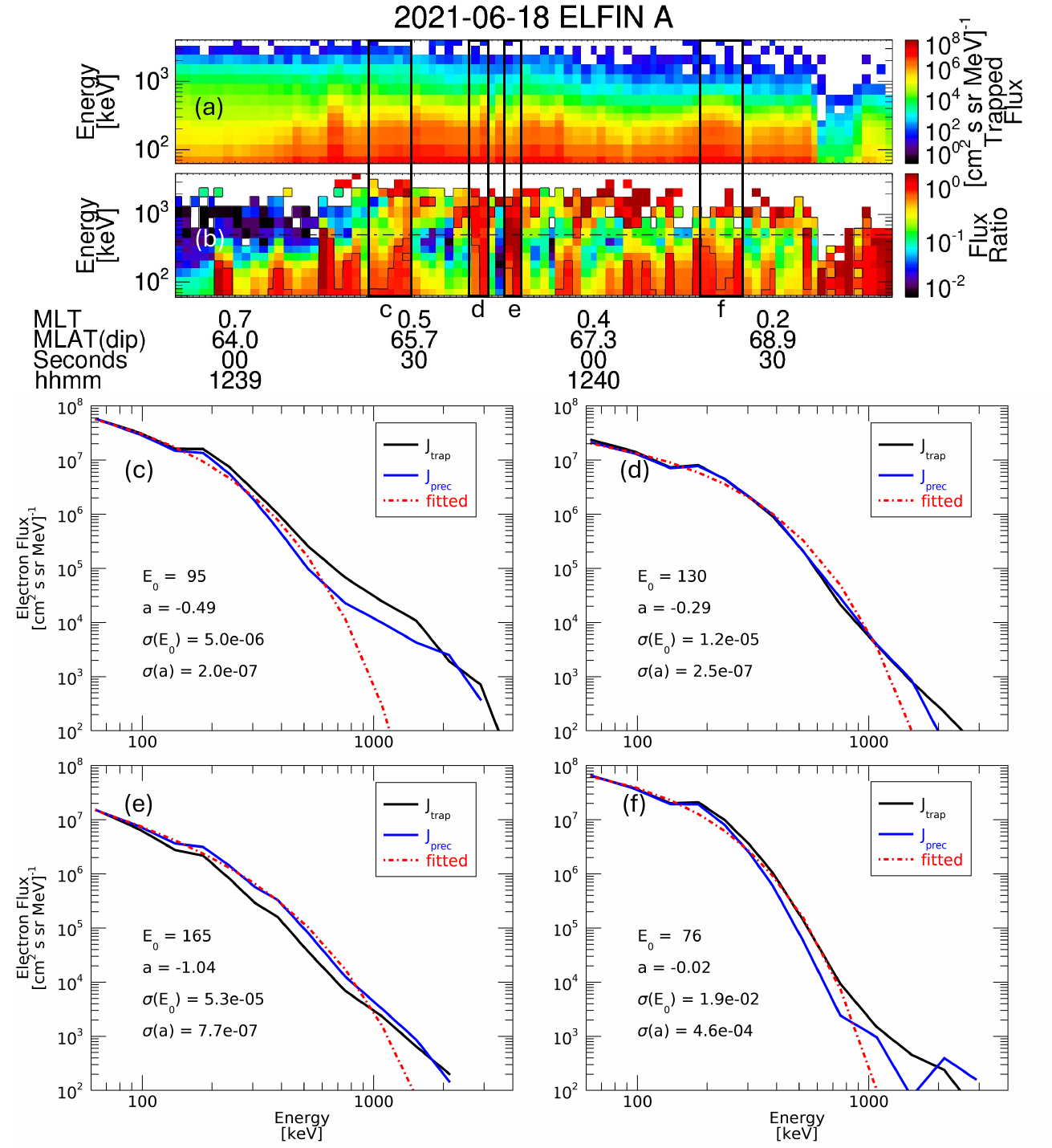}
\caption{Second example of night-side ELFIN observations of relativistic electron precipitation bursts from the outer radiation belt: (a) locally trapped electron fluxes, (b) precipitating-to-trapped flux ratio with precipitating bursts marked by black rectangles, (c)-(f) trapped and precipitating electron spectra within the relativistic bursts, with the precipitating spectrum fitted by $A\cdot\left(E/E_0\right)^a\cdot\exp(-E/E_0)$.              
\label{fig2}}
\end{figure*}

Figures \ref{fig1} and \ref{fig2} show two examples of ELFIN orbits from our night-side relativistic electron precipitation dataset. Panels (a) and (b) show the dynamic spectrum of trapped flux ($J_{trap}$)  and the ratio of precipitating flux and trapped flux ($J_{prec}/J_{trap}$). The plasma sheet region, characterized by isotropic fluxes ($J_{prec}/J_{trap}\sim 1$), is seen well in high magnetic latitudes. In the event shown in Figure \ref{fig1}, the plasma sheet at $MLAT>63^\circ$ \cite<fully isotropic flux region with the upper energy of electrons going down to higher latitudes; see discussion in>{Artemyev22:jgr:ELFIN&THEMIS}, with its inner edge marked by an electron isotropy boundary (IBe). Here, the dispersed precipitation pattern exhibits energies of isotropic fluxes increasing towards lower latitudes \cite<see discussions of formation of this pattern in>[and references therein]{Wilkins23,Artemyev23:ELFIN&dispersion,Sergeev23:elfin}. Equatorward of the IBe, ELFIN detects three bursts of intense precipitation with $J_{prec}/J_{trap}\sim 1$ at 50 keV and $J_{prec}/J_{trap}>0.1$ at 500 keV (indicated by the horizontal dashed line in panel (b)). For each burst, the average spectra of trapped and precipitating electrons are shown in panels (c), (d), and (e). The fluxes are quite similar, indicating the achievement of the strong diffusion limits on electron precipitation \cite{Kennel69}. The maximum energies where precipitating spectra are above the noise level ($\sim 100$ ${\rm cm}^{-2}{\rm s}^{-1}{\rm sr}^{-1}{\rm MeV}^{-1}$) reach up to $2$ MeV, which is comparable to the most energetic microburst precipitation \cite<see>{Douma17,Douma19,MeyerReed23}. The precipitating spectra are fitted with the function $const \cdot (E/E)^a\cdot\exp\left(-E/E_0\right)$ to quantify this relativistic precipitation, which combines power-law and exponential shapes \cite<see examples of this fitting for precipitating fluxes in>{Zhang22:microbursts}. For the first event, the typical energies of the exponential decay of precipitating fluxes are $E_0\sim 100-200$ keV, with $a$ around zero (Figure \ref{fig1}(c)-(e)). This suggests that there is no energy range with a power-law precipitation spectrum, and the entire precipitating flux follows the exponential spectrum of trapped electrons. The energy, $E_0\sim 100-200$ keV, indicates that relativistic precipitating fluxes (around $1-2$ MeV) are approximately $10^{-4}$ times smaller than fluxes of $50-100$ keV electrons, which are typically associated with whistler-mode wave driven precipitation \cite<e.g.,>{Li13:POES,Ni14:POES,Mourenas21:jgr:ELFIN,Tsai23}. $E_0$ can be slightly larger for trapped electrons because scattering at relativistic energies becomes less effective \cite{Summers07:rates}, and fluxes decay faster with increasing energy for precipitating electrons (compare the blue and black lines in panels (c)-(e)). Note that $J_{prec}$ being larger than $J_{trap}$ above $500$ keV is likely attributable to flux variability within the spin period of ELFIN rather than a physical effect, indicative of time aliasing due to the limited temporal resolution during rapidly varying dynamics \cite<see discussion in>{Zhang22:microbursts}.

Figure \ref{fig2} illustrates that whistler-mode driven precipitation overlap with those driven by EMIC waves (panels (a), (b)). Precipitation bursts related to whistler-mode waves show a decreasing $J_{prec}/J_{trap}$ ratio from sub-relativistic electron precipitation ($<500$ keV) to relativistic energies. However, these bursts overlap with $J_{prec}/J_{trap}\sim 1$ precipitation due to EMIC waves at energies $>1$ MeV \cite<see>[for additional examples of precipitation driven by a combination of whistler-mode and EMIC waves]{Bashir24:jgr}. The plasma sheet region, where electron fluxes are often below 300 keV with isotropic $J_{prec}/J_{trap}\sim 1$, is located at latitudes $>69^\circ$, while all precipitation bursts are observed equatorward of this region \cite<note that there is no clear IB for this event, a situation observed in roughly $20$\% of night-side ELFIN orbits, see discussion in>{Wilkins23,Artemyev24:jgr:ELFIN&IBe}. The fitting procedure effectively excludes the EMIC-driven precipitation spectrum and covers only the core precipitation below $500$ keV, which is driven by whistler-mode waves. Although we cannot fully distinguish precipitation drivers at intermediate energies ($500\sim 1000$ keV), where EMIC waves might cause electron losses through non-resonant or nonlinear wave-particle interactions \cite<see>{An22:prl,An24:jgr_EMICs,Grach&Demekhov23:theory,Hanzelka23:emic}, the fitting does not include purely EMIC-dominated losses at $>1$ MeV. The main difference between the precipitation bursts shown in Figures \ref{fig1}(c-e) and \ref{fig2}(c-f) is the $a$ parameter of the fitting: in Figure \ref{fig1}, $a\geq 0$ (with a plateau around $100-200$ keV), whereas in Figure \ref{fig2}, $a\leq 0$ (with rapid flux decreases as energy increases, even around $100-200$ keV). Both sets of bursts show strong diffusion limits ($J_{prec}/J_{trap}\sim 1$ at $<500$ keV) so the spectral differences ($a>0$ versus $a<0$) are attributed to variations in the trapped electrons.

Figures \ref{fig1} and \ref{fig2} have shown typical bursts of whistler-mode wave driven relativistic electron precipitation in the night-side inner magnetosphere (with average MLT values of 22 and 0.5 for these two ELFIN orbits). We compile statistics on such bursts and fit the precipitating spectra to investigate the dependencies of $E_0$ and $a$ on geomagnetic activity. For each burst, we also determine $E_{\max}$ as an energy where the level of flux goes down to $J_{prec}=100$ ${\rm cm}^{-2}{\rm s}^{-1}{\rm sr}^{-1}{\rm MeV}^{-1}$, i.e., $E_{\max}$ represents the maximum energy of precipitating fluxes above the noise level.

\section{Statistical properties of relativistic precipitation bursts}    
\begin{figure*}
\centering
\includegraphics[width=0.8\textwidth]{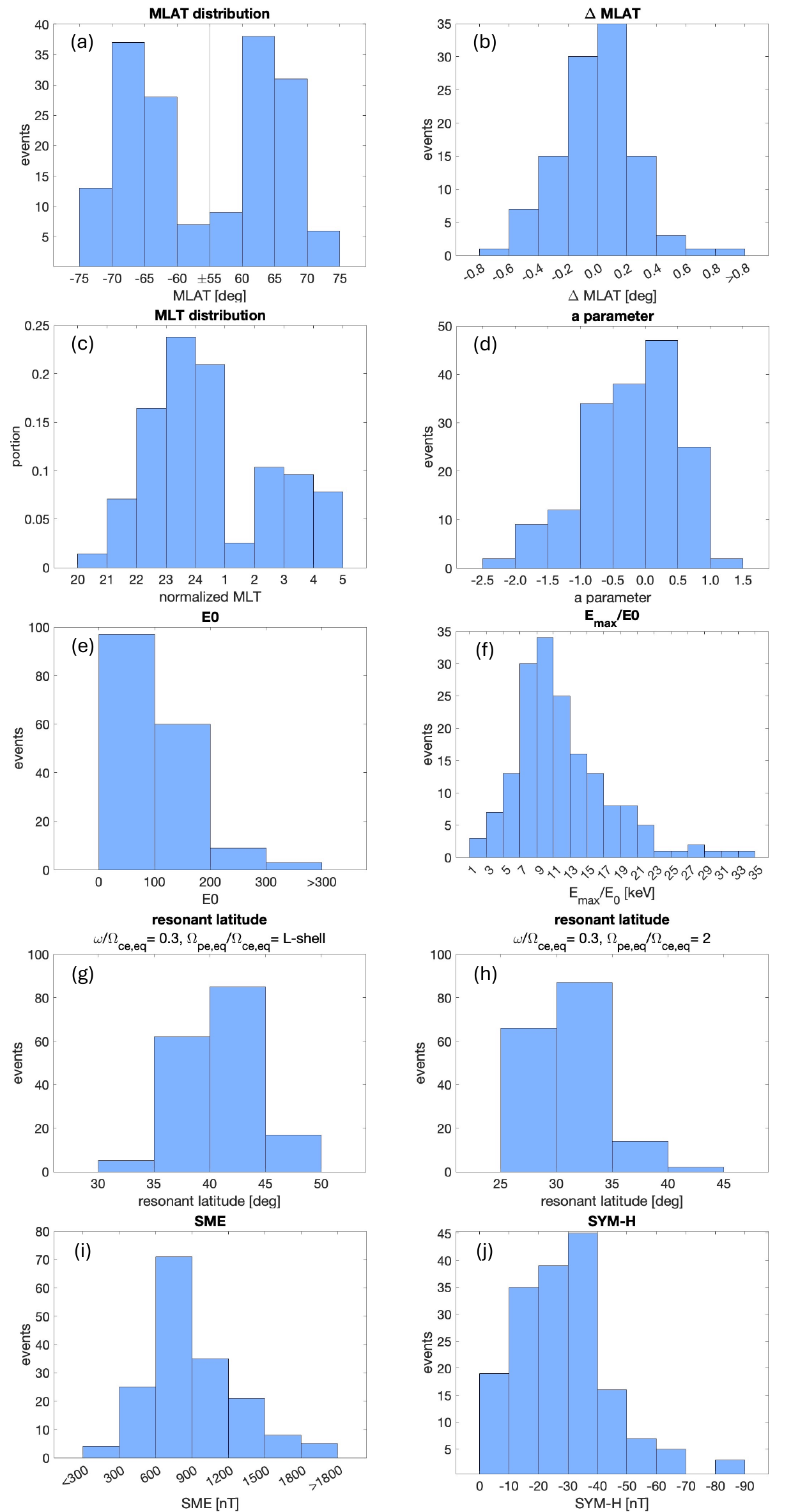}
\caption{Histograms with distributions of (a)-(b) $MLAT$ and $\Delta MLAT$, (c) $MLT$, (d)-(f) $a$, $E_0$ and $E_{max}/E_0$, (g)-(h) $MLAT_{res}$ for $\Omega_{pe}/\Omega_{ce}$ from the \cite{Sheeley01} model and for $\Omega_{pe}/\Omega_{ce}$=2 , (i)  $SME$ , and (j) $SYM$-$H$. In panel (c), the distribution is normalized by the total number of ELFIN orbits within each MLT bin \cite<see>[]{Tsai24:review}. 
\label{fig3}}
\end{figure*}

Figure \ref{fig3} shows the statistical characteristics of $169$ precipitation bursts with $E_{\max}>500$ keV. Almost all events are observed within the $[60,70^\circ]$ latitudinal range corresponded to the outer radiation belt in the night-side magnetosphere (panel (a)). The latitudinal range of these bursts is about $\Delta MLTA \leq 0.4^\circ$ (panel (b)), which corresponds to $\Delta r \leq 3000$ km when projected onto the equatorial plane \cite<see details of projections in the night-side in>{Artemyev24:jgr:ELFIN&IBe}. This spatial scale is typical for whistler-mode wave source regions associated with plasma sheet injections \cite{Agapitov17:grl,Zhang23:jgr:ELFIN&scales} as well as for whistler-mode chorus wave source regions in the outer radiation belt \cite<e.g.,>{Santolik&Gurnett03,Agapitov11:JGR}. Interestingly, most relativistic electron bursts are observed either in the pre-midnight sector (panel (c)), where most injections occur \cite{Gabrielse14}, or in the post-midnight sector, despite the latter being well known for whistler-mode wave activity \cite<e.g.,>{Agapitov18:jgr}. We suggest that relativistic electron precipitation requires not only wave activity, but also gradients in background plasma density and magnetic field associated with injections that can guide waves (see the Discussion section and \cite{Artemyev24:jgr:ELFIN&injection,Kang24:elfin}). Without such gradients, the waves generated by dawnward-drifting electrons \cite{Tao11} would be confined around the equator \cite{Meredith12} and scatter only sub-relativistic electrons. The importance of injections for relativistic electron precipitation is further supported by panels (i) and (j), which show that relativistic precipitation events are associated with strong geomagnetic activity. Most events occur with SME (an analog of the AE index) above $300$ nT \cite {Gjerloev12} during the substorm expansion/recovery phase associated with plasma sheet injections. Additionally, the majority of events happen during storms, meaning they are storm-time injections, with the SYM-H index smaller than $-30$ nT \cite<see discussion about the efficiency of storm-time injections and energetic particle transport in>{Beyene22,Beyene&Angelopoulos24}. Specifically, 146 out of 169 (86\%) events are associated with storms where the SYM-H minimum is below $-50$nT, with 51\% observed during the storm main phase and 49\% during the recovery phase. 

Typical energies from the fitting of electron precipitation spectra, $E_0$, are usually $<200$ keV (panel (e)), while the parameter $a$  is around zero or negative (panel (d)). This indicates that the precipitating electron fluxes decrease rapidly with increasing energy, with the flux ratio at $50$ keV being approximately $10^{4}$ larger than at $\sim 1-2$ MeV. The small absolute value of $a$ suggests that the precipitating spectra follow an exponential distribution. The average magnitude of the precipitating flux at $50$ keV is $A\approx 4.5\times10^{7}$ ${\rm cm}^{-2}{\rm s}^{-1}{\rm sr}^{-1}{\rm MeV}^{-1}$. Thus, the statistics from panels (d) and (e) can be used to define the range of relativistic night-side electron precipitation using the fitting function $A\cdot(E/E_0)^a\cdot\exp\left(1/2-E/E_0\right)$. Panel (f) shows the maximum energy of the precipitating electrons is about a factor of $10-20$ of $E_0$ for most events. Therefore, this fitting function should be applicable for energies $E<2$ MeV, assuming a typical $E_0\sim 100$ keV.

Using the resonance condition for field-aligned whistler-mode waves and the cold plasma dispersion relation \cite{bookLyons&Williams}, we evaluate the magnetic latitudes at which electron scattering occurs at $E_{\max}$ for our dataset of relativistic bursts \cite<see similar calculations in>{Roosnovo24:elfin,Artemyev24:jgr:ELFIN&injection}. One key parameter in determining the resonance latitude is the ratio of the plasma frequency to the electron cyclotron frequency, $\Omega_{pe,eq}/\Omega_{ce,eq}$, which depends on the plasma density.
We consider two scenarios: (1) $\Omega_{pe,eq}/\Omega_{ce,eq}\approx L$ from the model by Sheeley et al. (2001) \cite{Sheeley01} and (2) $\Omega_{pe,eq}/\Omega_{ce,eq}=2$, which aligns with observations of significant plasma density reduction during geomagnetically active periods, such as substorm injections \cite{Agapitov19:fpe}. The distribution of plasma density along magnetic field lines is based on the model by \cite{Denton06}. Panels (g) and (h) show the results of the resonance latitudes evaluated at the loss-cone pitch angle for precipitating electrons and $E_{\max}$. For both $\Omega_{pe,eq}/\Omega_{ce,eq}$ approximations, most events show scattering latitudes above $30^\circ$, which is significantly higher than the maximum latitudes of night-side whistler-mode waves \cite{Meredith12,Agapitov13:jgr}. Therefore, these relativistic precipitation bursts suggest the presence of a small population of ducted waves capable of propagating to middle latitudes ($\geq 30^\circ$) and scattering electrons there. 


\begin{figure*}
\centering
\includegraphics[width=1\textwidth]{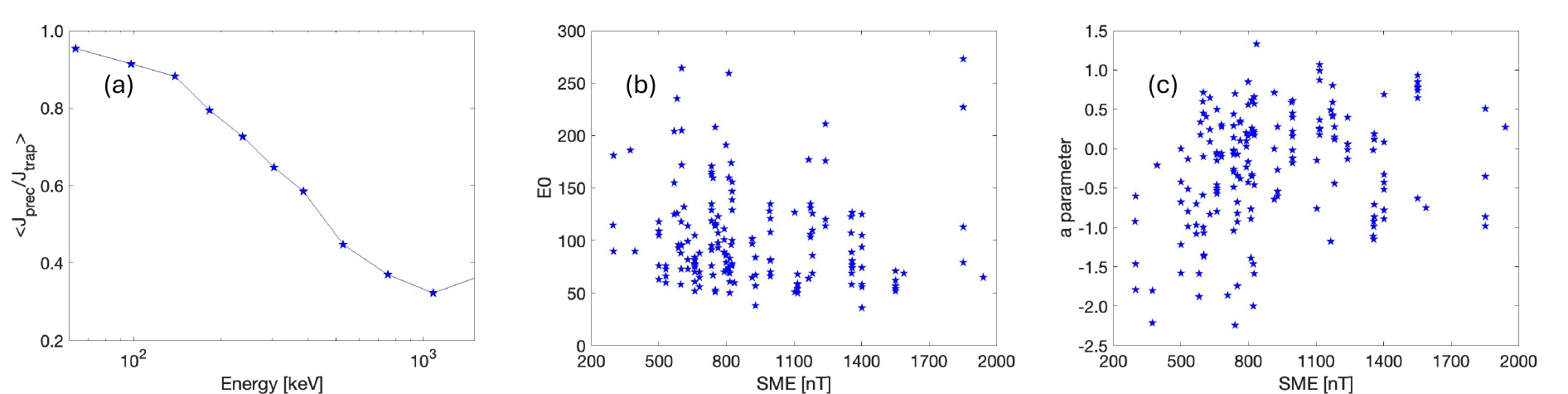}
\caption{Average spectrum of the precipitating-to-trapped electron flux ratio (a). Scatter plots showing the distributions of events in the  $(SME, E_0)$ space (b) , and in the $(SME, a)$ space (c).
\label{fig4}}
\end{figure*}

Figure \ref{fig4} (a) shows the average spectrum of the precipitating-to-trapped electron flux ratio for the entire dataset of events. Data points with $J_{prec}/J_{trap}>1.2$ were excluded because such measurements are associated with limitations of ELFIN measurements (see \citeA{Zhang22:microbursts}). The average flux ratio is close to the strong diffusion limit at $50$ keV ($\langle{J_{prec}/J_{trap}}\rangle\sim1$) and remains relatively high up to $1$ MeV ($\langle{J_{prec}/J_{trap}}\rangle\sim0.3$). This supports our initial hypothesis that night-side relativistic precipitation bursts are associated with extremely high wave activity and are likely driven by plasma sheet injections.

Figures \ref{fig4} (b) and \ref{fig4} (c) show the dependence of the fitting parameters $E_0$ and $a$ on geomagnetic activity. For weaker substorms, with $SME<300$ nT, the precipitation bursts exhibit $E_0>150$ keV and $a<-1$, indicative of precipitation with a power-law spectrum in the $[50,300]$ keV range \cite<see similar observations for microburst precipitation in>{Johnson21:FIREBIRD,Shumko23:FIREBIRD}, along with exponential decay above these energies. In contrast, the vast majority of precipitation bursts observed across all SME values in the $[300,1500]$ nT range are characterized by purely exponential spectra, with $E_0\leq100$ keV and $a\sim 0$. This suggests that stronger geomagnetic activity does not preferentially increase relativistic electron precipitation but instead maintains $E_0$ at $\sim100$keV and proportionately increases all energies accordingly while retaining the exponential term of the precipitation spectrum about $10^{-4}$ times smaller at $\sim 1-2$ MeV compared to $50-100$ keV.

\section{Discussion and concluding remarks}
\begin{figure*}
\centering
\includegraphics[width=1\textwidth]{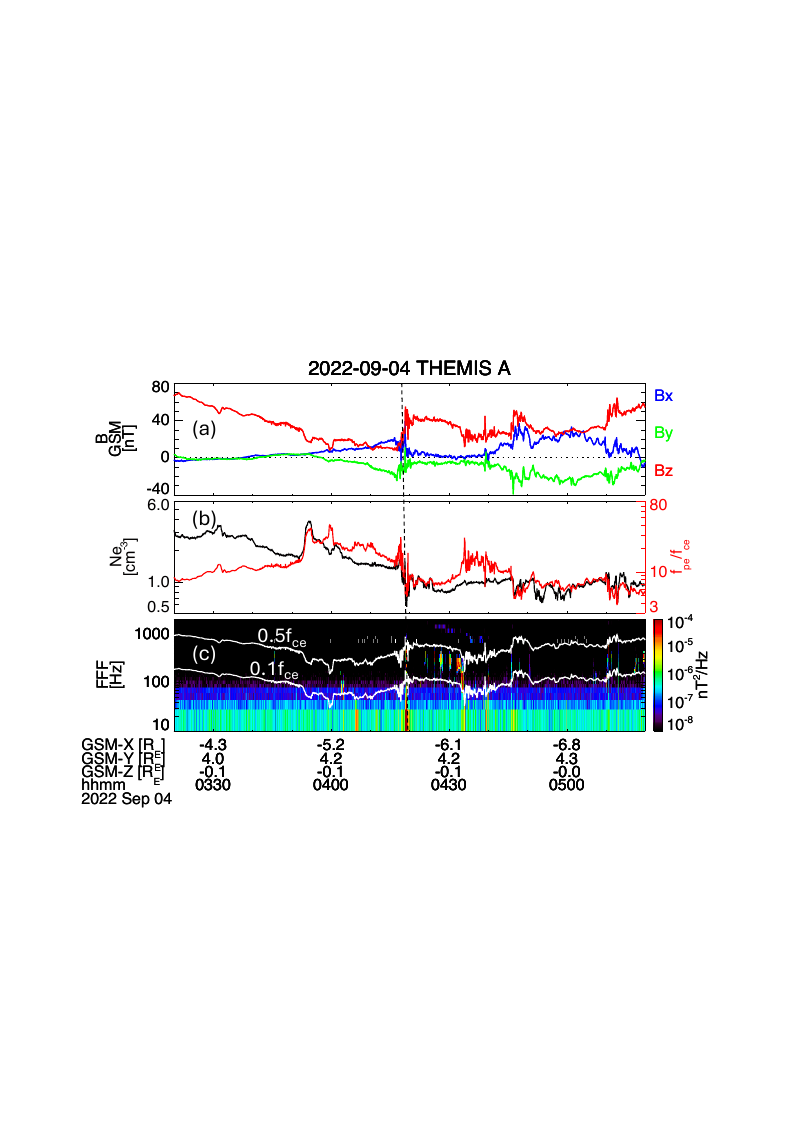}
\caption{ THEMIS A observations of the plasma injection conjugate to the ELFIN event from Figure \ref{fig1}: (a) magnetic field components in the GSM coordinates, (b) plasma density and the ratio of plasma frequency to electron gyrofrequency ($f_{pe}/f_{ce}$), and (c) the wave magnetic field spectra, with $f_{ce}/2$ and $f_{ce}/10$ shown by white lines. The vertical dashed line shows the approximate time of ELFIN's observations of relativistic precipitation bursts, which is close to the injection time at THEMIS A.
\label{fig5}}
\end{figure*}

\begin{figure*}
\centering
\includegraphics[width=1\textwidth]{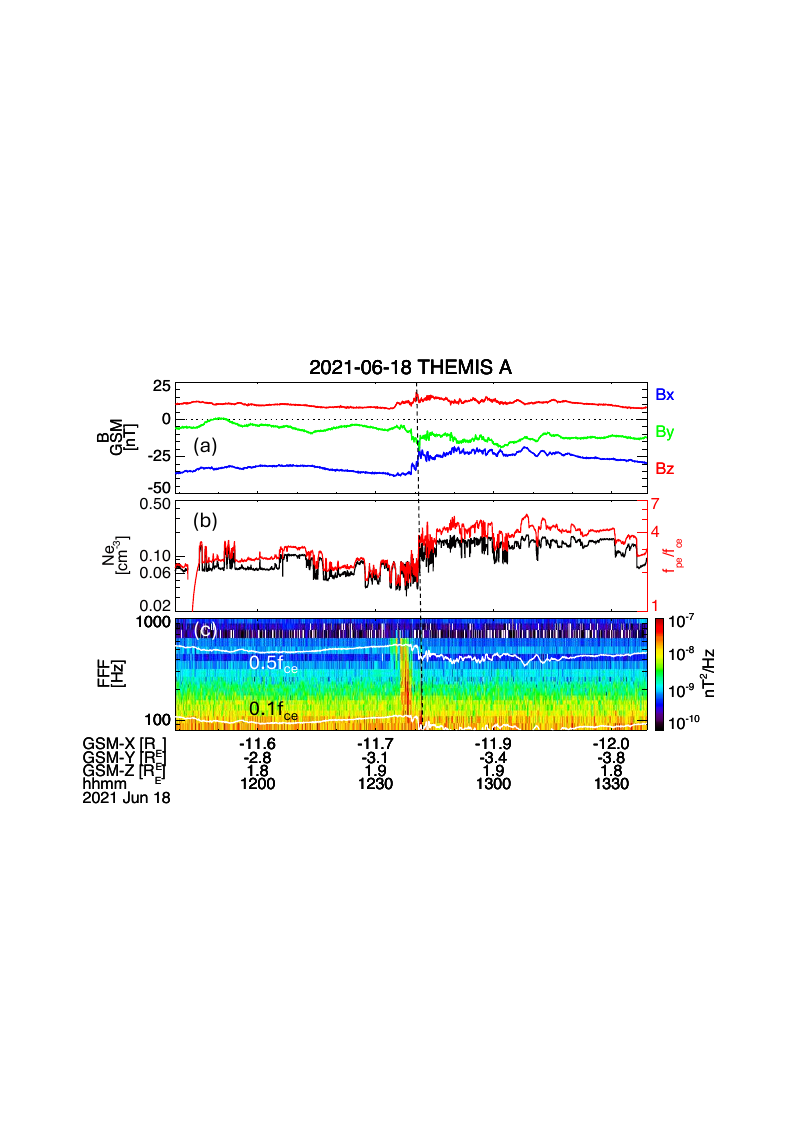}
\caption{  THEMIS A observations of the plasma injection conjugate to the ELFIN event from Figure \ref{fig2}: (a) magnetic field components in the GSM coordinates, (b) plasma density and the ratio of plasma frequency to electron gyrofrequency ($f_{pe}/f_{ce}$), and the(c) wave magnetic field spectra, with $f_{ce}/2$ and $f_{ce}/10$ shown by white lines. The vertical dashed line shows the approximate time of ELFIN's observations of relativistic precipitation bursts, which is close to the injection time at THEMIS A.
\label{fig6}}
\end{figure*}

In this study, we investigated relativistic electron bursts observed by the low-altitude (low-Earth orbit) ELFIN CubeSats in the night-side inner magnetosphere. All burst occur equatorward of the inner edge of the electron plasma sheet and are very likely associated with electron scattering by whistler-mode waves. The precipitating-to-trapped flux ratio at $50$ keV, with $\langle{J_{prec}/J_{trap}}\rangle\sim1$, indicates that these bursts can be generated by intense waves \cite<above $50$pT; see>{Tsai22}. The high ratio at $1$ MeV suggests that these waves are likely ducted, propagating from the equatorial source region to magnetic latitudes greater than $35^\circ$ without significant damping \cite{Chen22:microbursts,Kang24:elfin}. We attribute the generation of very intense whistler-mode waves and their ducting to the same phenomena associated with plasma sheet injections. 
\cite<see also discussion in>{Artemyev24:jgr:ELFIN&injection}. These injections can transport anisotropic electron populations \cite<e.g.,>{Zhang18:whistlers&injections,Ukhorskiy22:NatSR} and lead to the generation of the most intense whistler-mode waves \cite<e.g.,>{LeContel09,Deng10,Tao11,Fu14:radiation_belts,Grigorenko20:whistlers}. Additionally, plasma density gradients at injection fronts (dipolarization fronts) \cite{Runov11jgr,Hwang11} may contribute to wave ducting. To support this hypothesis, we examine two events from our statistics (two ELFIN orbits) for which THEMIS provides near-equatorial measurements in the same MLT sector as the ELFIN observations of relativistic electron precipitation. Figures \ref{fig5} and  \ref{fig6} show coincident THEMIS observations during the ELFIN events in Figures \ref{fig1} and \ref{fig2}.

The comparison between THEMIS and ELFIN is most representative of the event shown in Figures \ref{fig1} and \ref{fig5}. THEMIS A, located at $MLT\sim 21$ and $L\sim 7$, observes a plasma sheet injection at 04:20 UT. A clear magnetic field dipolarization, characterized by an increase in $B_z$ (panel (a)), occurs at the injection front (dipolarization front) \cite{Nakamura04,Runov09grl}. This front is associated with a strong plasma density gradient (panel (b)) and a decrease in the ratio of the plasma frequency to the gyro frequency. After the front (from 04:20 to 04:30 UT), THEMIS A observes multiple plasma density fluctuations and structured whistler-mode wave emissions (panel (c)). Whistler-mode waves are likely generated by hot anisotropic electrons, which are consistently observed around the dipolarization front  \cite<e.g.,>{Fu13:NatPh,Runov13,Zhang19:grl:whistlers,Breuillard16}. The density fluctuations and the strong gradient at the front can facilitate the trapping and ducting of these waves \cite<see discussions in>{Hanzelka&Santolik19,Streltsov&Bengtson20,Streltsov&Goyal21}. Around the same time (04:20 UT), ELFIN A captures a series of relativistic precipitation bursts at $MLT\sim 22$, one hour from THEMIS A (see Figure \ref{fig1}). ELFIN's observations are equatorward of the IBe, meaning these precipitation bursts may be projected closer to Earth than the injection observed by THEMIS \cite<see statistics on the IBe location in>{Wilkins23}. Therefore, a plausible scenario involves an injection propagating earthward, with whistler-mode wave sources and density fluctuations facilitating wave ducting to the region of ELFIN observations \cite<see one such example in>{Artemyev22:jgr:DF&ELFIN}.

Figures \ref{fig2} and \ref{fig6} illustrate similar ELFIN/THEMIS observations of relativistic electron precipitation bursts around 12:40 UT at $MLT\sim0.5$ and a plasma sheet injection at $MLT\sim1$. THEMIS A, located below the equatorial plane, shows $B_x\sim -25$ nT and $B_z\sim15$ nT (panel (a)), and thus may be projected further down the tail along magnetic field lines from its location at approximately $12R_E$. Similar to the event in Figures \ref{fig1} and \ref{fig5}: THEMIS A observes an injection before it reaches the inner magnetosphere, which brings in whistler-mode wave sources associated with the relativistic electron precipitation observed by ELFIN. A strong plasma density gradient at the injection front (detected as a decrease in $|B_x|$ and an increase in $B_z$; see details in \cite{Nakamura04,Runov09grl}) and density fluctuations around the front are observed (panels (a) and (b)). Moreover, slightly ahead of the front, THEMIS A detects whistler-mode waves (panel (c)), characterized by peaks of wave activity around half of the electron gyrofrequency. These waves differ from the broad-band low-frequency turbulence, which is likely associated with kinetic Alfven waves \cite{Chaston12,Malaspina18}. Therefore, the near-equatorial observations of THEMIS A (Figures \ref{fig5} and \ref{fig6}) support our hypothesis that plasma sheet injections likely drive relativistic electron precipitation in the night-side magnetosphere.
These injections transport whistler-mode sources and density gradients responsible for wave ducting to middle latitudes.

In conclusion, this study investigates relativistic electron precipitation on the night side through a statistical analysis of ELFIN observations. We show that:
\begin{itemize}
  \item These precipitation bursts are linked to intense near-equatorial whistler-mode waves, which account for $J_{prec}/J_{trap}\sim 1$ at $50$ keV and $J_{prec}/J_{trap}\sim 0.3$ at $1$ MeV as they propagate to middle latitudes (with the resonance latitude $\sim 35^\circ$).
  \item The precipitating spectrum in the energy range of $<1-2$ MeV can be approximated by $A\cdot(E/E_0)^a\cdot\exp\left(1/2-E/E_0\right)$, with $A\approx 4.5\times 10^{7}{\rm cm}^{-2}{\rm s}^{-1}{\rm sr}^{-1}{\rm MeV}^{-1}$, $E_0\approx 100$ keV, and $a\in[-0.5,0.5]$. 
  \item These precipitation bursts are associated with substorm activity ($SME>300$ nT) and are likely driven by plasma sheet injections during the substorm expansion phase.
  \item The spatial extent of relativistic precipitation at $MLT\in[22,1]$ and $MLAT\in[60^\circ,70^\circ]$ is consistent with the idea that these bursts are related to substorm injections and are distinct from typical whistler-driven precipitation observed in the post-midnight sector.
\end{itemize}
These results suggest that relativistic precipitation bursts should be treated as a distinct precipitation band, which may have been overlooked in global models of radiation belt dynamics. Current models may not fully capture this phenomenon, which are often based on average whistler-mode wave statistics with strong wave intensity constraints around the equator \cite{Meredith12,Agapitov13:jgr}. The primary driver of these precipitation events is likely ducted whistler-mode waves, a wave population that has been under-investigated \cite<see discussion in>{Artemyev24:jgr:ELFIN&injection,Kang24:elfin}. These night-side relativistic electron losses are significant for understanding radiation belt dynamics and may impact magnetosphere-ionosphere coupling at subauroral latitudes. These precipitation events could disturb the ionosphere below $100$ km, influencing various ionospheric processes \cite<see discussions in>{Oyama17:eep,Nishimura21:agu,Yu22:ssr,Zou24}.

\acknowledgments
X.J.Z., A.V.A., and X.L. acknowledge support by NASA awards 80NSSC23K0108, 80NSSC23K1038, 80NSSC23K0100, 80NSSC24K0561, 80NSSC24K0138, 80NSSC19K0844, and NSF grant 2329897.

We are grateful to NASA’s CubeSat Launch Initiative for ELFIN's successful launch in the desired orbits. We acknowledge early support of ELFIN project by the AFOSR, under its University Nanosat Program, UNP-8 project, contract FA9453-12-D-0285, and by the California Space Grant program. We acknowledge critical contributions of numerous volunteer ELFIN team student members. We acknowledge NASA contract NAS5-02099 for the use of data from the THEMIS Mission.

\section*{Open Research}
\noindent Data was retrieved and analyzed using PySPEDAS and SPEDAS, see \citeA{Angelopoulos19}.



\end{document}